\documentclass[reprint,aps,prb]{revtex4-1}
\usepackage{graphicx,epstopdf,float,setspace,mathtools,amsmath,multirow,physics,xcolor}
\pdfoutput=1
\begin{document}
\title{Landauer's Principle in a Quantum Szilard Engine Without Maxwell's Demon}
\author{Alhun Aydin$^{1,3}$}
\author{Altug Sisman$^{2,3}$}
\author{Ronnie Kosloff$^{1}$}
\email{ronnie@fh.huji.ac.il}
\affiliation{$^{1}$Fritz Haber Research Center, Institute of Chemistry, The Hebrew University of Jerusalem, 91904, Jerusalem, Israel \\
$^{2}$Department of Physics and Astronomy, 75120, Uppsala University, Uppsala, Sweden \\
$^{3}$Nano Energy Research Group, Energy Institute, Istanbul Technical University, 34469, Istanbul, Turkey}
\date{\today}
\begin{abstract}
Quantum Szilard engine constitutes an adequate interplay of thermodynamics, information theory and quantum mechanics. Szilard engines are in general operated by a Maxwell's Demon where Landauer's principle resolves the apparent paradoxes. Here we propose a Szilard engine setup without featuring an explicit Maxwell's demon. In a demonless Szilard engine, the acquisition of which-side information is not required, but erasure and the related heat dissipation still take place implicitly by the very nature of the work extraction process. We see that the insertion of the partition in a quantum Szilard engine does not localize the particle to one side, instead it creates a superposition state of the particle being in both sides. To be able to extract work from the system, particle has to be localized at one side. The localization occurs as a result of quantum measurement on the particle, which shows the importance of the measurement process regardless of whether one uses the acquired information or not. In accordance with the Landauer's principle, localization by quantum measurement corresponds to a logically irreversible operation and for this reason it has to be accompanied by the corresponding heat dissipation. This shows the validity of the Landauer's principle even in quantum Szilard engines without Maxwell's demon. Furthermore, we take quantum confinement effects fully into account to analyze the Szilard cycle in the quantum regime thoroughly and obtain highly accurate analytical expressions for work and heat exchanges. Our results show that Landauer's principle holds the key role to understand the thermodynamics of the localization of the particle by quantum measurement, which explicitly saves the second law in demonless engines and shows that quantum-mechanical considerations are essential to reconcile thermodynamics and information theory.
\end{abstract}
\maketitle
\section{Introduction} 
The second law of thermodynamics has been challenged by thought experiments many times. While the second law eludes itself from all charges so far, the interrogators have caused to reveal subtle links and features between different disciplines of physics. One of these thought experiments, quantum Szilard engine, holds the key for a possible reconciliation of thermodynamics, information theory and quantum mechanics in the same footing. Understanding the full role of Landauer's principle in a quantum Szilard engine along with its restrictions and liberations is still a crucial challenge in the emerging field of quantum information thermodynamics\cite{review1}.

The first link between thermodynamics and information has been unraveled by Maxwell in 1871\cite{OMaxwell}. To briefly summarize his thought experiment, consider a container with two equally sized compartments where each compartment is filled with a gas at thermal equilibrium having the same pressure and temperature. An information processing being (later coined as the Maxwell's demon) controls a tiny opening; closes if a molecule comes towards left compartment and opens if it comes towards right, thereby creating a pressure (or temperature if it does the selection according to molecules' energy) difference from a thermal equilibrium. Without any expenditure of work, the demon transfers heat from colder region to hotter region and decreases the entropy, which presents a clear violation of the Clausius statement of the second law of thermodynamics. 
Instead of an information processing demon, Smoluchowski proposed a trap-door model in 1912\cite{OSmoluchowski}, but these kind of Brownian ratchet-like devices cannot operate reliably at thermal equilibrium because of random fluctuations.
Szilard took the problem one step further by designing a heat engine that appears to violate the second law\cite{OSzilard}. Szilard's classical heat engine can be described as follows: Consider a container with a single gas molecule inside and in contact with a heat reservoir. Now insert a piston having zero thickness at the middle of the container which splits it into two parts with equal volumes. Depending on the position of the molecule one can attach a weight to the piston with a string over a pulley, which makes it possible to extract work by the expansion of a freely movable piston caused by the pressure that molecule exerts. To return the initial state, the partition can be removed without any work consumption and the whole process can be repeated in a cyclic manner. All thermodynamic processes are defined as isothermal and reversible. This engine apparently violates the Kelvin-Planck statement of the second law (that is actually equivalent to the Clausius statement) by converting heat directly into equivalent amount of work through a cyclic process.

Information-theoretic arguments for the resolution of Maxwell's demon and Szilard engine paradoxes put forward explicitly by von Neumann, Brillouin and Rothstein during the beginning of 50's\cite{stanford}. However, the major leap forward on the issue came in 1961 by Landauer who stated that any logically irreversible operation must be accompanied by some heat dissipation, which is called Landauer's principle now\cite{OLandauer}. In this context, Landauer gave an explicit example of resetting a memory to the reference state, which is nowadays called as information erasure. Later by Penrose and independently by Bennett, this has been used to resolve the Szilard problem. Penrose and Bennett argued that in order to complete the cycle in Szilard engine, one needs to erase the information from the demon's memory (by resetting) which generates a heat dissipation\cite{OPenrose,OBennett}. After Penrose and Bennett, information erasure was the key to resolve the problem and the measurement was thought to cost no work. Indeed, for classical measurements this is the case\cite{OPenrose}. Nowadays, for the resolution of the classical versions of both Maxwell's demon and Szilard engine paradoxes, it has been accepted by most that the demon must store the information about the molecule(s) and need to erase it to complete the cycle\cite{infoerasure,revmodphysvedral}. 

Although Szilard's thought experiment does not mention about quantum mechanics, by the very nature of the problem, it is clear that such an engine needs to operate at nanoscale where quantum effects may play a significant role. Zurek\cite{zurek86,physrepzurek} was one of the first to explicitly take into account the role of quantum superposition in a quantum Szilard engine. However, it has been first realized by Biedenharn and Solem\cite{BS} that quantum mechanics might be essential for the resolution of the paradox. They also pointed out the resemblance between Szilard engine problem and double-slit experiment. The role of wavefunction collapse during a measurement in the perspective of Maxwell's demon has been also discussed\cite{bender}.
Through an Ising model adaptation of the Szilard engine, Parrondo\cite{symbreak,symbreak2} argued that symmetry breaking is the key ingredient in the operation of a Szilard engine. Recently, Alicki and Horodecki proposed a postulate to separate ergodicity from the quantum superpositions, \textit{i.e.} the non-existence of the superpositions of pure quantum states belonging to different ergodic components, which is relevant to understand the thermodynamics of the Szilard engine\cite{alicki}.
During the last decades, quantum versions of Maxwell's demon and Szilard engine have been explored by considering specific physical systems as well as abstract ones\cite{seth,PhysRevLett.106.070401,particlestat,PhysRevLett.120.100601,PhysRevE.83.061108,revisitquantum,PhysRevE.85.031114,PhysRevLett.121.030604,qusziwithoutheat}. Experimental realization claims of the Maxwell's demon and Szilard engine has been made also during the last decade\cite{experiment2010,experiment2012,experiment2014,PhysRevLett.115.260602,photonian,experiment2017,PhysRevLett.120.210601}.

Quantum Szilard engine is crucial in the sense that it constitutes the backbone of the reconciliation of thermodynamics, information and quantum theories. Therefore, the coverage of all of its aspects and a fully-acclaimed resolution of the problem are still required. Many arguments have been appeared and circulated in literature based on some assumptions and logic flows, but to our knowledge, there is still no in depth investigation and simulation of the quantum version of the original Szilard engine under quasistatic isothermal processes and confinement effects.

In this work, we start with a new Szilard engine setup without the Maxwell's demon, and argue that acquisition of which-side information and feedback control may not be essential in a Szilard engine. Without actually recording and storing the information, the explicit necessity of the erasure is discussed. We present by various novel examples that even in the absence of Maxwell's demon, erasure might take place implicitly. The consequences of implicit erasure are also addressed in Sec. II. In the light of demonless classical Szilard engines, we argue that the demonless resolution of Szilard's paradox implies the requirement of quantum mechanical considerations. In Sec. III, we examine the thermodynamic cycle of a quantum Szilard engine with quasistatic isothermal processes at all steps. Confinement effects are fully taken into account and they help to understand and analyze the cycle thoroughly. We show that despite the lack of need for acquiring the which-side information, the localization of particle at one side by quantum measurement is still essential for the operation of a Szilard engine, which brings out the necessity of quantum-mechanical considerations. On the contrary case, the engine won't work due to the superposition of the particle being at both sides. The quantum measurement process is therefore crucial in a quantum Szilard engine, even in the absence of an explicit information processing. Localization by quantum measurement is a logically irreversible process and just like the information erasure, it has to be accompanied by a corresponding heat dissipation. Consequently, in a demonless Szilard engine, quantum mechanics validates the Landauer's principle. We calculate free energy, entropy and internal energy changes in each step of the quantum Szilard engine by taking quantum confinement effects fully into account. Due to quantum size and shape effects, insertion requires a work to be done, though it is exactly recovered back during the expansion process along with $kT\ln 2$. Detailed numerical simulations also supports our arguments. Highly accurate analytical expressions for work and heat exchanges in the quantum regime of a Szilard engine cycle are provided in the appendix. We summarize our findings in Sec. IV and conclude by mentioning the possible future extensions of this work.

\section{A Szilard engine without Maxwell's Demon?} 
The thermodynamic process of a basic memory erasure is described in Refs.\cite{pleniovitelli,revmodphysvedral} by simply reversing the Szilard engine cycle. In literature, Szilard engines have been almost always designed with an external memory device of this kind implemented by a Maxwell's Demon through some protocols. Landauer's erasure explicitly takes place on the memory device. On the other hand, several demonless Szilard engine setups have also been proposed before to demonstrate a complete thermodynamic cycle without the which-side information acquisition, such as the trap-door model by Smoluchowski\cite{OSmoluchowski}, a mechanical setup by Popper\cite{stanford}, an automatic device by Jauch and Baron\cite{JB} and recently by Alicki\cite{alicki}. Some demonless engines, like Smoluchowski's trap-door, won't work because of thermal fluctuations. The final stance of the other setups is still open to debate\cite{stanford,HS,norton,ENorton1,ENorton2,elecdemon}.

Here, we address this issue again and by presenting a new demonless Szilard engine setup (Fig. 1) we discuss the implications and consequences by means of Landauer's principle. In our proposed setup, a partition (piston) divides the Szilard's box into two equally sized compartments and magnetic rods are attached to the piston on both sides of the box along with solenoids. After the insertion of the piston, regardless of the position of particle (either left or right classically), the piston will cause the magnetic rods to move within the solenoid and generate a positive and/or negative electric current by electromagnetic induction. This current can then be converted into a direct current, if required, by a passive diode bridge to drive a DC motor. Even without using a rectifier component like a diode bridge, the current can still be utilized to drive an AC motor by designing the opposite winding directions in each solenoid. Even repositioning of the piston won't cause any problem, as long as we disconnect the load from the bridge during the repositioning process of the piston. With this setup, work can be extracted from the Szilard engine, without having the necessity of knowing which side of the container the particle is. Since no information acquisition has occurred, there would be no need for a memory to record the which-side information and so the cycle seems to be completed without any apparent erasure of information. This device has the sole effect of absorbing heat from a reservoir and converting it to the equivalent amount of work, thus, seems to violate the Clausius' version of 2nd law!

\begin{figure}[t]
\centering
\includegraphics[width=0.48\textwidth]{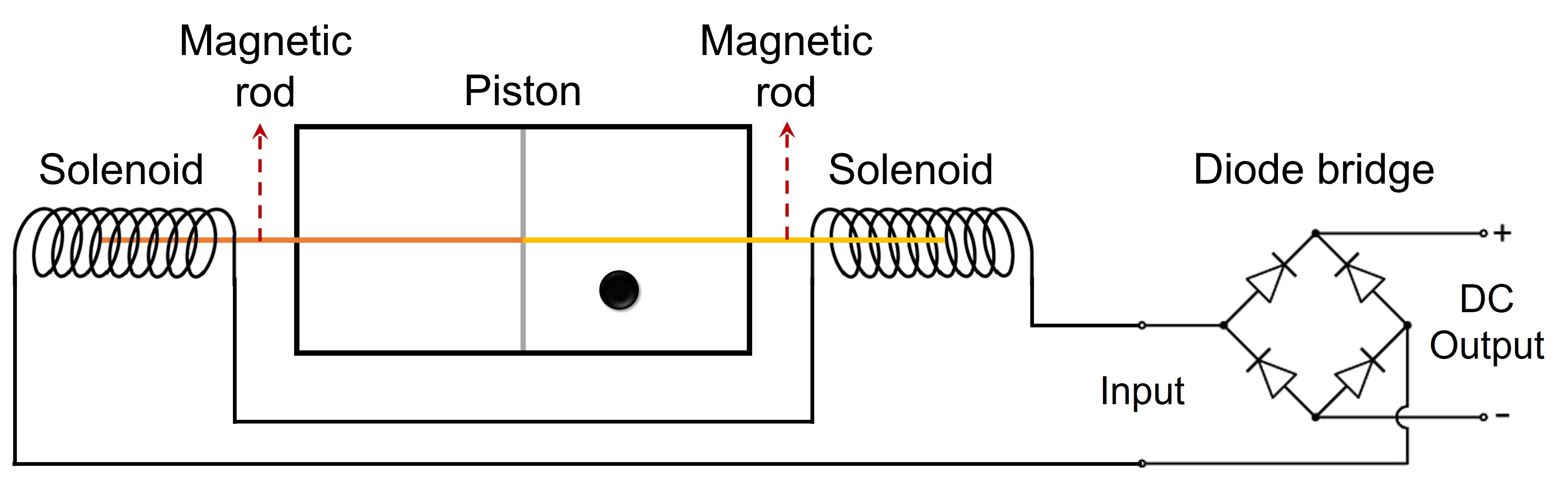}
\caption{A classical Szilard engine setup without Maxwell's demon. Schematic shows the step after the partition (piston) symmetrically inserted. Magnetic rods are attached to the both sides of the piston and two solenoids are placed on both ends which are connected to a passive diode bridge. Regardless of the particle's position, magnetic rods moving inside the solenoids induce an electric current that can be utilized from the output of the diode bridge.}
\label{fig:pic1}
\end{figure}

One can object to this erasure-free interpretation by stating that in any demonless engine, there would be some mechanism that implicitly stores the which-side information and corresponds to a some sort of memory which needs to be erased eventually\cite{Magnasco,LeffRex1,LeffRex2,szidemrevisit}. For instance, in our proposed setup, left and right solenoids induce the electric current independent of the position of the particle inside the box. However, this state still encodes the information about which outcome occurs, though we are not necessarily using it to extract work. Converting this undirected current to a direct current by a diode bridge might correspond to the erasure of this which-side information and if that's the case, this must be accompanied by a heat dissipation. Even in the case of the absence of a rectifier diode bridge, the positive or negative phase information will be erased irreversibly by the internal physical mechanisms of an AC motor. In every possible case, two-way motion (leftward/rightward or clockwise/counterclockwise) is always converted into a one-way motion to be able to make use of it as a thermodynamic work. By the very nature of this conversion, information loss happens and it is impossible to recover the input information from the output. This means by the Landauer's principle, there has to be an entropy generation and corresponding heat dissipation. These kind of attempts to bypass Landauer's principle is similar to the attempts that were done in the past to bypass the second law of thermodynamics, which all eventually have been shown to be impossible. There seems no way to bypass the Landauer's principle, even by constructing Szilard engines without Maxwell's demon. In the following section, we explore the quantum version of a demonless Szilard engine and extend the usage of Landauer's principle beyond the memory erasure.

\section{Thermodynamics of a demonless quantum Szilard engine under confinement effects} 

In this section, we revisit the conceptual Szilard engine by considering quantum-mechanical effects like quantum measurement, localization and confinement (both size and shape) effects. We keep our discussion as close as it can to the original Szilard engine, albeit it is quantum-mechanical version of it. It is hard to make general statements out of Szilard engine setups with very specific physical system considerations. Staying on thermodynamic footing allows us to make more generalized statements. Therefore, to make our arguments model-independent, we design our setup without any concerns of a physical realization. Schematic of a single-particle quantum Szilard heat engine is presented in Fig. 2. The setup is composed of three components, namely the system $\mathcal{S}$, measuring device $\mathcal{D}$ and heat bath $\mathcal{B}$ at temperature $T$. The thermodynamic cycle consists of four steps; insertion (I), measurement (II), expansion (III) and removal (IV). The partition has zero thickness and symmetrically divides the container into two compartments. All boundaries confining the particle are perfectly impenetrable. $\mathcal{B}$ is in contact with $\mathcal{S}$ and $\mathcal{D}$ all the time to keep all processes isothermal. To stick with the original propositions and to prevent any unnecessary complications of the problem, we assume all processes to be quasistatic. 

\begin{figure}[t]
\centering
\includegraphics[width=0.48\textwidth]{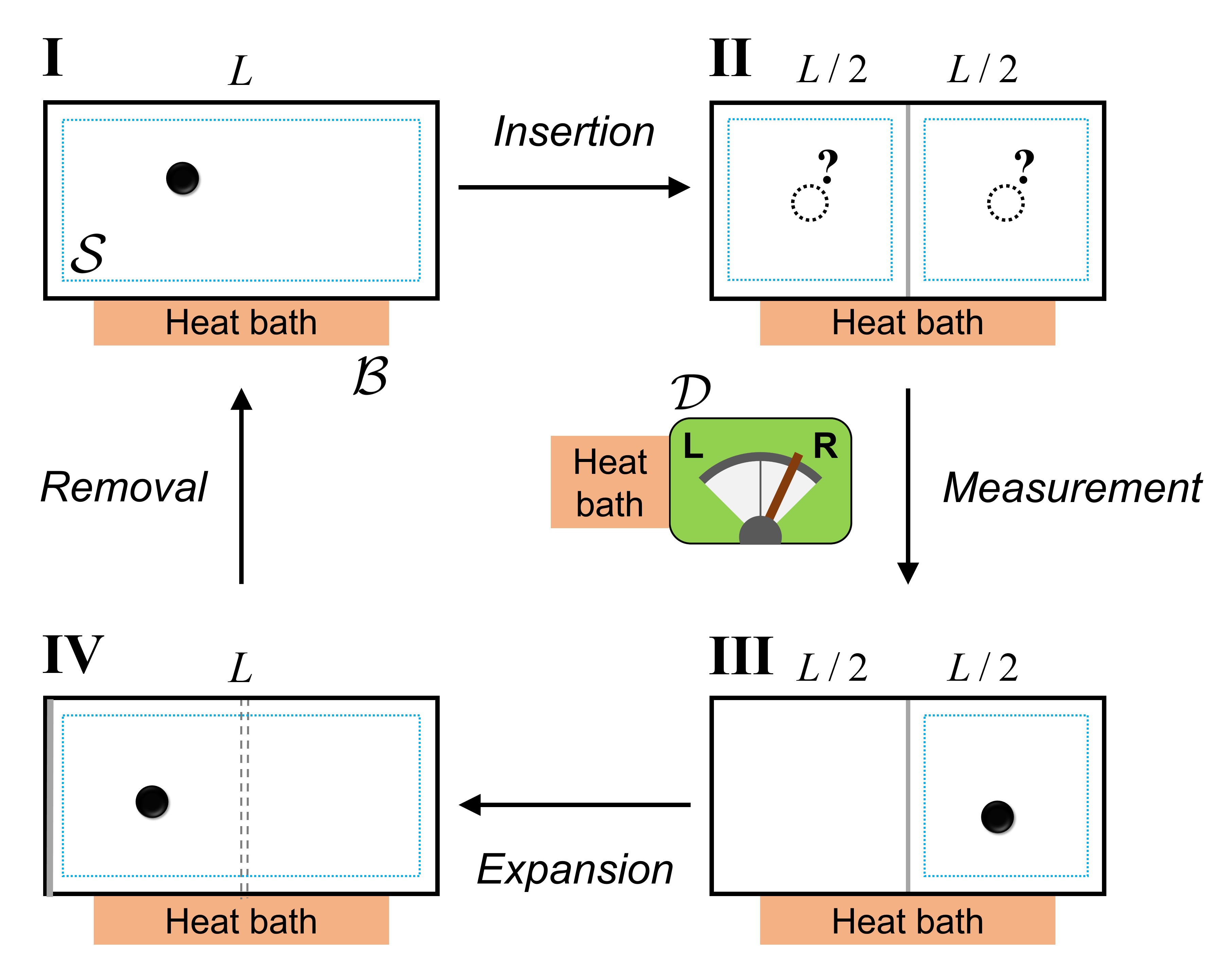}
\caption{A quantum Szilard engine setup composed of three components, system $\mathcal{S}$, measuring device $\mathcal{D}$ and heat bath $\mathcal{B}$. $\mathcal{S}$ denotes the container with a single particle inside. $\mathcal{D}$ is the device which measures the particle. $\mathcal{B}$ at temperature $T$ is in contact with both $\mathcal{S}$ and $\mathcal{D}$, keeping all processes isothermal. Dotted turquoise lines denote the effective regions that the particle occupies because of the confinement effects. (I$\rightarrow$II) Inserting the partition into the container. Symmetric insertion divides the container into two equal compartments and creates an entangled state of the particle's position. (II$\rightarrow$III) Performing quantum measurement to localize the particle into one of the compartments. (III$\rightarrow $IV) Letting particle to expand the partition and extracting work from the system. (IV$\rightarrow $I) Removing the partition from the container at the boundary, which completes the cycle.}
\label{fig:pic2}
\end{figure}

For a 1-dimensional Szilard box, initial wavefunction of the confined particle before the insertion of the partition is given by $\bra{x}\ket{\psi_n}=\sqrt{2/L}\sin\left(n\pi x/L\right)$, where $n$ is quantum state variable. The density matrix completely describing the equilibrium state of the single particle in a quantum Szilard engine reads
\begin{equation}
\rho=Z^{-1}\sum_n \exp\left(-\beta E_n\right)\ket{\psi_n}\bra{\psi_n},
\end{equation}
where $\beta=1/(k T)$ is inverse temperature with $k$ denotes the Boltzmann's constant and $T$ temperature. Here, $Z$ is the single particle partition function which is given by
\begin{equation}
Z=\sum_{n} \exp\left(-\beta E_n\right),
\end{equation}
where $E_n=n^2 h^2/(8mL^2)$ are 1D energy eigenvalues from the solution of Schr\"{o}dinger equation for the particle with mass $m$ inside the container with length $L$ at the initial stage. We choose bare electron mass for the calculations. Note that since there is only one particle inside the box, Maxwell-Boltzmann statistics is used as the distribution function.

By using the partition function, Helmholtz free energy is written as
\begin{equation}
F=-kT \ln Z.
\end{equation}
Entropy of the quantum thermodynamic system is written by von Neumann entropy as
\begin{equation}
S=-k \text{tr}\left(\rho\ln\rho\right),
\end{equation}
and then, internal energy of the system is as follows
\begin{equation}
U=kT Z^{-1}\sum_n E_n \exp\left(-\beta E_n\right).
\end{equation}

In the following subsections, we investigate the steps of the thermodynamic cycle and analyze work and heat exchanges between $\mathcal{S}$, $\mathcal{D}$ and $\mathcal{B}$ in detail for each step by evaluating free energies, entropies and internal energies of the thermodynamic states of the system. For the sake of clarity of the thermodynamic processes, we examine a Szilard engine with a 2-dimensional (2D) container in particular, rather than 1D. It should be noted that, although free energy, entropy and internal energy values differ in 2D and 1D cases, differences in these quantities (so work and heat exchanges) are the same for both cases, because momentum eigenvalues parallel to the direction of the inserted partition do not change during thermodynamic processes. The contributions from the direction parallel to the inserted partition cancels out when one takes the differences of thermodynamic quantities. Hence, work, heat and energy exchanges presented in this work are universal in the sense that they are independent from how many dimensional space the container is considered in.

\subsection{Step I: Creating superposition by insertion}

In the first step, a partition with zero thickness is quasistatically inserted into the system which is in contact with the heat bath $\mathcal{B}$ at temperature $T$. Although the insertion process is not a trivial step in a quantum Szilard engine, it is not thoroughly investigated before. What happens to the quantum particle in a quasistatic insertion along with the full thermodynamic analysis of the process are presented in Fig. 3 in detail. To make also a quantitative analysis, we choose a container with sizes $L_x=20$nm, $L_y=10$nm and temperature $T=300$K.

\begin{figure}[t]
\centering
\includegraphics[width=0.48\textwidth]{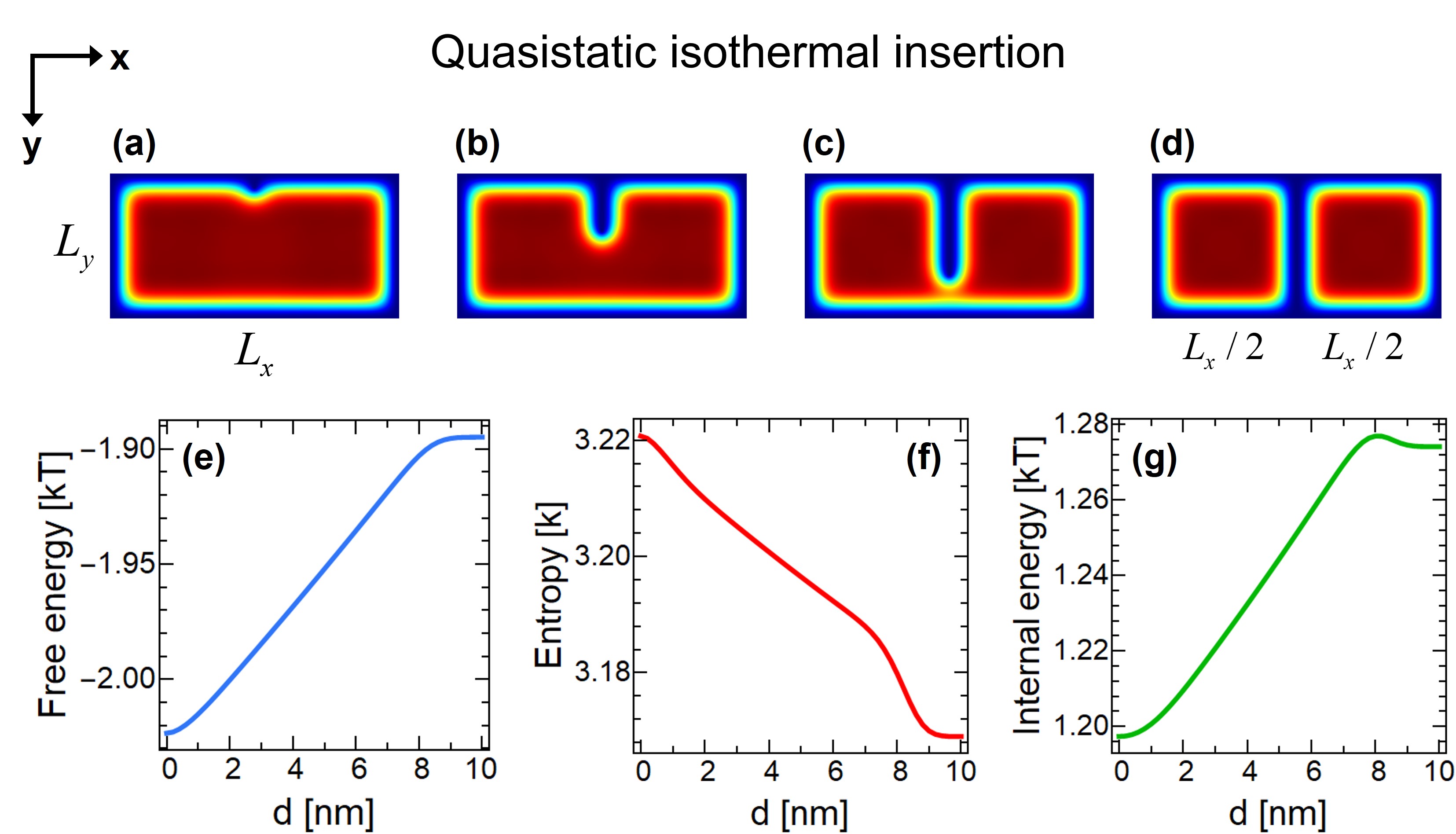}
\caption{Simulation of quasistatic isothermal insertion process for a container with sizes $L_x=20$nm, $L_y=10$nm and at temperature $T=300$K. $d$ denotes the depth of the partition inserted into the domain. (a) Quantum-mechanical thermal probability density distribution of the particle is non-uniform inside the container due to quantum size effects. Magnitudes of the density distributions are represented by the rainbow color scale, where red and blue colors denote higher and lower density regions respectively. Partition having zero thickness entering the container in $y$-direction, $d=1$nm. (b) Partition enters almost halfway of the container, $d=4$nm. Although it has zero thickness, confined particle perceives a finite effective thickness ($2\delta$). (c) Partition is at $d=7$nm depth. (d) Partition separates the container into two equally sized compartments, $d=10$nm. Particle has equal probability to be at both sides. Variation of (e) Helmholtz free energy, (f) entropy and (g) internal energy with respect to partition's penetration depth $d$ in nm's.}
\label{fig:pic3}
\end{figure}

A single quantum particle at thermal equilibrium occupies the whole domain in an inhomogeneous way due to its wave nature. The quantum-mechanical thermal probability density distribution of the confined particle is given by
\begin{equation}
n(\textbf{r})=Z^{-1}\sum_n \exp\left(-\beta E_n\right)\left|\Psi_n(\textbf{r})\right|^2,
\end{equation}
where $\textbf{r}$ is the position vector. Note that Eq. (6) contains both the thermal and quantum probabilities and so capturing quantum-thermodynamic nature of the confined system.

In Fig. 3a, 3b, 3c and 3d, numerical simulations\cite{comsol} of the density distributions of the confined particle is shown for insertion depths of $d=1$nm, $d=4$nm, $d=7$nm and $d=10$nm, respectively. $d$ denotes the depth of the inserted partition in $y$-direction. Density at a particular point inside the domain is represented by the rainbow color scale, where red and blue colors respectively denote higher and lower densities. As is seen, although partition dives into the domain, there is still possibility for particle to travel inbetween compartments, until the partition finally closes the opening at $d=10$nm, in Fig. 3d.

It is crucial to notice that, even after the partition divides the domain into two, particle has probability to occupy both regions of left and right compartments, as particle does not teleport itself into one of the compartments at the moment of the perfect division. This can be seen quantitatively and visually from the plots of Eq. (6) in Fig. 3. In fact, at the closure of the partition, particle becomes in a quantum superposition state of being both left and right sides at the same time. In other words, the insertion of the partition creates an entangled position-state of the particle. Entanglement is provided by the spatial correlation of states. Similar to the entangled photons in a double-slit experiment, the confined particle in a quantum Szilard engine is in a quantum state that is composed of possibilities occupying left and right compartments at the same time with equal probability in the symmetric insertion case. In this sense, insertion process in a quantum Szilard engine can be seen analogous to the entanglement creation of a photon by a beam splitter.

Density distributions of confined systems and variation in thermodynamic properties can be suitably understood and interpreted by the quantum boundary layer (QBL) concept\cite{qbl}. In the thermodynamics of confined systems, when particles are confined in nanoscale domains at thermal equilibrium, because of their wave nature, a layer with less occupation probability is formed near to boundaries, which is called QBL. In Fig. 3(a, b, c and d), the effect of QBL formation can clearly be seen. The particle in a quantum Szilard engine occupies effectively a smaller volume than the apparent geometric volume (the effective region is denoted by dotted turquoise regions in Fig. 2). Although inserted partition has zero thickness, the quantum particle confined in the domain perceives an effective thickness ($2\delta$) of the partition, which in turn changes the thermodynamic properties of the system, unlike in classical case. The concept of QBL not only gives a physical understanding to the quantum thermodynamics of confined systems, but also gives possibility to obtain analytical expressions of thermodynamic quantities (given in Appendix) with a quite high accuracy.

Free energy of the system increases during the insertion process, Fig. 3e, suggesting that it requires work to be done. Free energy slowly changes during the initial entrance of the partition, because of the less thermal probability density between $d=0$ and $d=1$nm due to QBL near to container's boundaries. Then, free energy linearly increases during the insertion until around $d=8$nm where QBLs of partition and the container wall starts to overlap\cite{qshe} in which free energy saturates to its final value at $d=10$nm. Free energy variation is approximately equal to the effective pressure times the effective thickness which corresponds to the quantum force exerted on the partition with zero thickness\cite{qforce}.

Free energy of the system before insertion is given by $F_{\text{I}}=-kT\ln[Z(L)]$. Partition function has also temperature and $L_y$ dependencies of course, but we won't denote them in the expressions, since they are constant during all processes. Likewise, $L_x$ is shortened to $L$ for brevity in expressions. Since insertion divides the system into entangled states occupying two compartments, the partition function of the final system after the insertion is the sum of two partition functions of left and right compartments. Hence, free energy after insertion is $F_{\text{II}}=-kT\ln[2Z(L/2)]=-kT\ln[Z(L/2)]-kT\ln 2$. It can be seen that $kT\ln 2$ term naturally arises inside the free energy expression $F_{\text{II}}$, as a consequence of the quantum entanglement of the particle which is now in both compartments at once. From the difference of initial and final free energies, insertion work reads
\begin{equation}
\begin{split}
W_{\text{ins}}&=kT\ln\frac{Z(L)}{2Z(L/2)} \\
&=kT\ln\frac{Z(L)}{Z(L/2)}-kT\ln 2.
\end{split}
\end{equation}
Due to quantum size effects, the result of the above equation is non-zero. If one takes the limit of $L\rightarrow\infty$ or $T\rightarrow\infty$, insertion work will be zero. The quantitative examination giving the functinoal behavior of this limit is shown in Fig. 4 by black and gray curves for size and temperature behaviors respectively. They go to zero in the infinite volume and temperature limits, but blow up to infinity in $L\rightarrow 0$ or $T\rightarrow 0$ limit, because of the exponential in the partition function. 

During the insertion, entropy of the system decreases as the available effective volume that can be occupied by the particle decreases, because of QBLs on both sides of the inserted wall. Entropy of the system before insertion is $S_{\text{I}}=-k \text{tr}[\rho(L)\ln\rho(L)]$ and after it becomes $S_{\text{II}}=-k \text{tr}[\rho(L/2)\ln\rho(L/2)]+k\ln 2$. Then, from the difference of initial and final entropies, heat dissipation to the environment during the insertion process becomes
\begin{equation}
\begin{split}
Q_{\text{ins}}=kT\ln 2+kT\text{tr}\left[\rho(L)\ln\rho(L)-\rho(L/2)\ln\rho(L/2)\right].
\end{split}
\end{equation}
Variation of $Q_{\text{ins}}$ with domain size and temperature is examined in Fig. 4a and 4b by purple and pink curves respectively. For large domain sizes and high temperatures it goes to zero. However, when domain size becomes extremely small, say less than around 10nm's in our case, it changes its negatively increasing behavior and rapidly goes up to the positive amplitude. Although this particular confinement effect is also interesting and it is actually related to quantum shape effects\cite{qshe}, it will be investigated in another study and it won't change the arguments for the resolution of the Szilard engine problem in this work. 

Internal energies of initial and final stages are respectively written as $U_{\text{I}}=kT Z^{-1}(L)\sum_n E_n(L)\exp[-\beta E_n(L)]$ and $U_{\text{II}}=kT Z^{-1}(L/2)\sum_n E_n(L/2)\exp[-\beta E_n(L/2)]$. Therefore, change in internal energy of the system in insertion process becomes
\begin{equation}
\begin{split}
\Delta U_{\text{ins}}=&kTZ^{-1}\left(\frac{L}{2}\right)\sum_n E_n\left(\frac{L}{2}\right)\exp[-\beta E_n\left(\frac{L}{2}\right)] \\
&-kTZ^{-1}(L)\sum_n E_n(L)\exp[-\beta E_n(L)].
\end{split}
\end{equation}
During the insertion process, $\mathcal{S}$ interacts only with $\mathcal{B}$ and all work and heat exchanges happen between them.

Although in an isothermal process internal energy change is zero, this statement is valid only classically. Due to quantum size effects, internal energy (as well as other thermodynamic properties) does not only the function of temperature, but also the domain sizes\cite{qse}. Variations of changes in internal energy with respect to $L$ and $T$ are examined in Fig. 4a and 4b by the teal and turquoise curves respectively.

Analytical expressions for work and heat exchanges, as well as changes in internal energy, are given in Appendix. They perfectly match with the expressions given by Eqs. (7), (8) and (9), that's why we didn't plot them in Fig. 4. The relative errors are less than $10^{-6}$ even for $L=10$nm at $T=300$K. For even stronger confinements, one needs to consider quantum shape effects\cite{qshe} as well in order to get even more accurate representations, but for our analysis here, the analytical expressions given in Appendix are quite sufficient. 

\begin{figure}[t]
\centering
\includegraphics[width=0.48\textwidth]{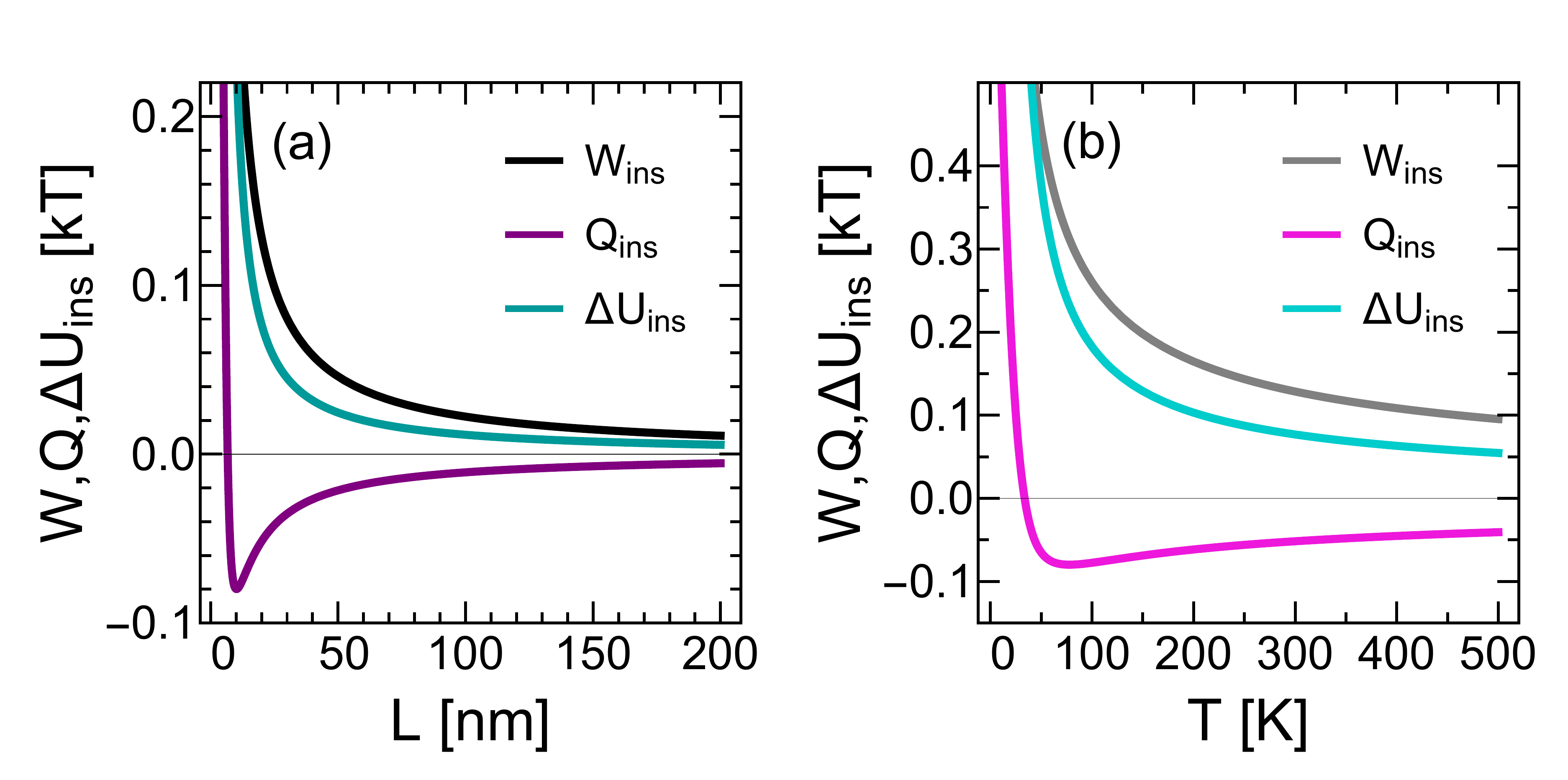}
\caption{Work and heat exchanges as well as changes in internal energy during insertion process with respect to (a) domain size $L$ and (b) temperature $T$. Black/gray, teal/turquoise and purple/pink curves represent insertion work, insertion heat and internal energy change during insertion between $\mathcal{S}$ and $\mathcal{B}$ respectively for (a)/(b).}
\label{fig:pic4}
\end{figure}

\subsection{Step II: Localization by quantum measurement}

In the beginning of step II, we have the superposition of the particle in left and right compartments of the container. In this stage, we argue that it is impossible to extract work from the Szilard engine unless we localize the particle into one of the compartments. To demonstrate this, we analyze the system in quasistatic isothermal expansion without the localization of the particle in Fig. 5. $l$ denotes the position of the partition in $x$-direction. First of all, piston won't move because particle is in both left and right sides with equal probability and therefore exerting equal but opposite pressure to the piston, Fig. 5a. This fact can also be seen from the free energy variation in Fig. 5e with respect to the location of the piston (at $l=10$nm piston is at the center, at $l=20$nm it's on the right boundary of container). Free energy stays almost constant until around 3 QBL thickness left between the piston and domain boundaries, which is due to the so called quantum shape effect\cite{qshe}. Behavior of entropy and internal energy in Fig. 5f and 5g after $l=7$nm can also be explained by quantum shape effects. The discussion of these behaviors is beyond the scope of this article as it won't change the presented arguments and only has quantitative effect, however one can check Ref. \cite{qshe} for the explanations of these behaviors via similar systems. What Fig. 5 shows us is, the best we can do in a quantum Szilard engine without localization is to get back the work that we expend during the insertion process. This can be seen by comparing the free energy differences of both processes shown in Figs. 3e and 5e, where the differences in free energies are exactly equal to each other and $kT\ln 2$ term does not pop out unlike in the localized case. Therefore, without the localization of the particle, net work output during the cycle will be zero. One cannot extract work in this case, because the partition won't move as it feels equal amount of pressure from both compartments, see Fig. 5a and 5e. Consequently, although we don't need to acquire the which-side information, we still have to measure the particle in order to localize it. Here, we are not strictly interested in the physical structure of the measuring device. Its only function is to localize the wavefunction of the particle and it has ability to receive heat from $\mathcal{B}$.

\begin{figure}[t]
\centering
\includegraphics[width=0.48\textwidth]{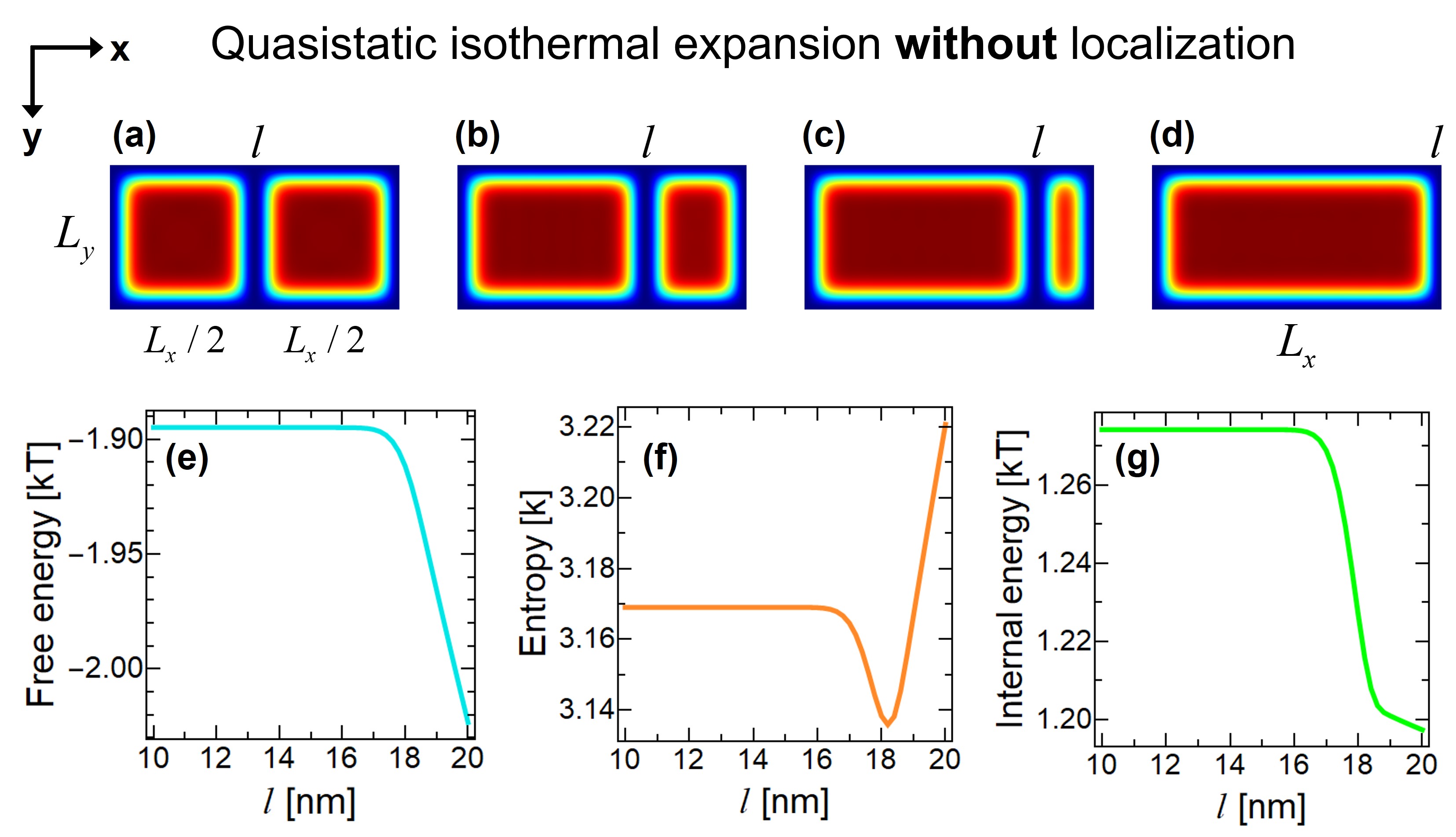}
\caption{Simulation of quasistatic isothermal expansion process without performing the quantum measurement and localizing the particle. $l$ denotes the lateral position of the partition. (a) Partition symmetrically divides the container into two equally sized compartments. (b) Partition is moved to the position of $l=14$nm by preserving quantum superposition and without localizing the particle. (c) Partition is at $l=17$nm. (d) Partition is moved to the right boundary of the container, $l=20$nm. Initial situation before the insertion is recovered. Variation of (e) Helmholtz free energy, (f) entropy and (g) internal energy with respect to partition's lateral position $l$ in nm's.}
\label{fig:pic5}
\end{figure}

In literature, measurement has been thought to cost no work (and corresponding heat dissipation). This is indeed true since gaining (or writing) information is just a mapping operation which preserves the phase space volume and therefore can be realized without heat dissipation, in principle. However, the operation of localizing the particle by quantum measurement does not preserve the phase space volume. It is not possible to infer the initial state from the final state, since there is no one-to-one correspondence between the pre-measurement and post-measurement states. Although measurement process is thermodynamically reversible, it has a logical irreversibility and by Landauer's principle it is necessarily dissipative\cite{OLandauer,OBennett}. Logical irreversibility does not imply thermodynamic irreversibility\cite{thermoinfo} and localization by quantum measurement constitutes an example of this. Classical measurement is logically reversible (Landauer-Bennett) but quantum measurement is not. In literature, it is somewhat mistakenly assumed as if the erasure is the only way for a logical irreversibility. However, this is not necessarily true. Resetting a memory erases any trace regarding the initial trace of the information, however, localization by quantum measurement does also the exactly same operation in the context of logical irreversibility.

Information that is acquired should be independent from whether we use it or not and quantum measurement exactly does that. Independent from the usage of the information, the particle localized at a one definite side. After the localization, storing of the information can be done without any expenditure of work.

Before the localization, the particle was in both boxes at once and hence the total partition function of the system was the sum of the partition functions of the two compartments which is $2Z(L/2)$. After the localization particle finds itself in one of the boxes and partition function reduces to $Z(L/2)$. After localization, free energy becomes $F_{\text{III}}=-kT\ln[Z(L/2)]$. This suggest that free energy change in the system is $kT\ln 2$. Hence, measurement process causes a work input to the system which is given by, 
\begin{equation}
W_{\text{msr}}=kT\ln 2.
\end{equation}
Similarly, after the localization of the particle, entropy of the system becomes $S_{\text{III}}=-k \text{tr}[\rho(L/2)\ln\rho(L/2)]$. We know also the entropy of the system at step II. By taking their difference, the heat dissipated is found as
\begin{equation}
Q_{\text{msr}}=-kT\ln 2.
\end{equation}
which is exactly the same amount of the work that is expended during the localization by measurement process. Internal energy of the system does not change during this process ($U_{\text{II}}=U_{\text{III}}$) so that
\begin{equation}
\Delta U_{\text{msr}}=0.
\end{equation}
During the measurement process, $\mathcal{S}$ interacts with $\mathcal{D}$, so work and heat exchanges happen between them through $\mathcal{B}$. It should be noted that $\mathcal{D}$ does not exactly correspond to the Maxwell's demon, because it does not function as a memory device, its only function is quantum measurement and collapsing the particle's wavefunction.

Since we consider $\mathcal{S}$ and $\mathcal{D}$ separately, the heat dissipation appears during the measurement process which causes localization. Several authors have also argued that this actually does not contradict with Landauer's principle and the link between thermodynamics and information\cite{PhysRevA.80.012322,PhysRevE.86.040106,thermoinfo}. They claim that this discrepancy is just a matter of interpretation, however, the thermodynamic processes of the erasure-free interpretation have never been shown explicitly before in the context of quantum Szilard engine, although work cost of the operations related with the entropy change and information has also been shown\cite{PhysRevE.79.041102,PhysRevLett.102.250602,PhysRevE.84.061110,natcominfo,PhysRevE.91.032118,PhysRevX.3.041003,newjphysinfo,PhysRevLett.111.230402,PhysRevE.94.010103,physrep,PhysRevX.8.021011}.

Work cost of measurement has also been taken into account from purely information-theoretic point of view in literature\cite{PhysRevE.86.040106,PhysRevE.84.061110,PhysRevLett.111.230402,thermoinfo,PhysRevE.94.010103}. In this context, extractable work from the system has mutual information term in addition to the free energy difference in the system, so that $W_{\text{ext}}^{\mathcal{S}}\leq-\Delta F^{\mathcal{S}}+kTI$ where $I$ is the mutual information between $\mathcal{S}$ and $\mathcal{D}$, which is $I=\ln 2$ in this case. The mutual information term corresponds to the work cost of measurement which is also equivalent to the net work due to entanglement \cite{workgain}. Since measurement on the system increases its free energy by $kT\ln 2$, extractable work becomes zero in a thermodynamically reversible cyclic process\cite{thermoinfo}. Hence, our localization by measurement picture is also compatible with the information-theoretic picture with mutual information.

\subsection{Step III: Work extraction by expansion}

After localizing the particle, now it's possible to extract work from the system by isothermal expansion. In Fig. 6, simulation of the expansion process is shown in several steps. Particle starts with the localized volume of $L/2$ (Fig. 6a), expands (Fig. 6b, 6c) and the piston reaches to the end of the container (Fig. 6d). This expansion can be converted into mechanical work by simple mechanisms.

The system consists only of half of the compartment in the beginning of step III, and consequently the partition function is Z(L/2) and free energy is written as $F_{\text{III}}=-kT\ln[Z(L/2)]$. Entropy of the system becomes $S_{\text{III}}=-k \text{tr}[\rho(L/2)\ln\rho(L/2)]$. In Fig. 6e, 6f and 6g, changes in free energy, entropy and internal energy of the system during the expansion are plotted. While entropy of the system is increasing, free energy and internal energy decrease during the expansion process, which is an expected result. Expansion work can be written as
\begin{equation}
\begin{split}
W_{\text{exp}}=&kT\ln\frac{Z(L/2)}{Z(L)} \\
&=-W_{\text{ins}}-kT\ln 2,
\end{split}
\end{equation}
and heat exchange between the $\mathcal{S}$ and $\mathcal{B}$ is 
\begin{equation}
\begin{split}
Q_{\text{exp}}=&kT\text{tr}\left[\rho(L/2)\ln\rho(L/2)-\rho(L)\ln\rho(L)\right] \\
&=-Q_{\text{ins}}+kT\ln 2.
\end{split}
\end{equation}
Internal energy change of $\mathcal{S}$ during the expansion process is non-zero and negative and it is equal to $\Delta U_{\text{ins}}$ with the opposite sign,
\begin{equation}
\begin{split}
\Delta U_{\text{exp}}=-\Delta U_{\text{ins}}.
\end{split}
\end{equation}
During the expansion process, $\mathcal{S}$ only interacts with $\mathcal{B}$ and all work and heat exchanges occur between them.

\begin{figure}[t]
\centering
\includegraphics[width=0.48\textwidth]{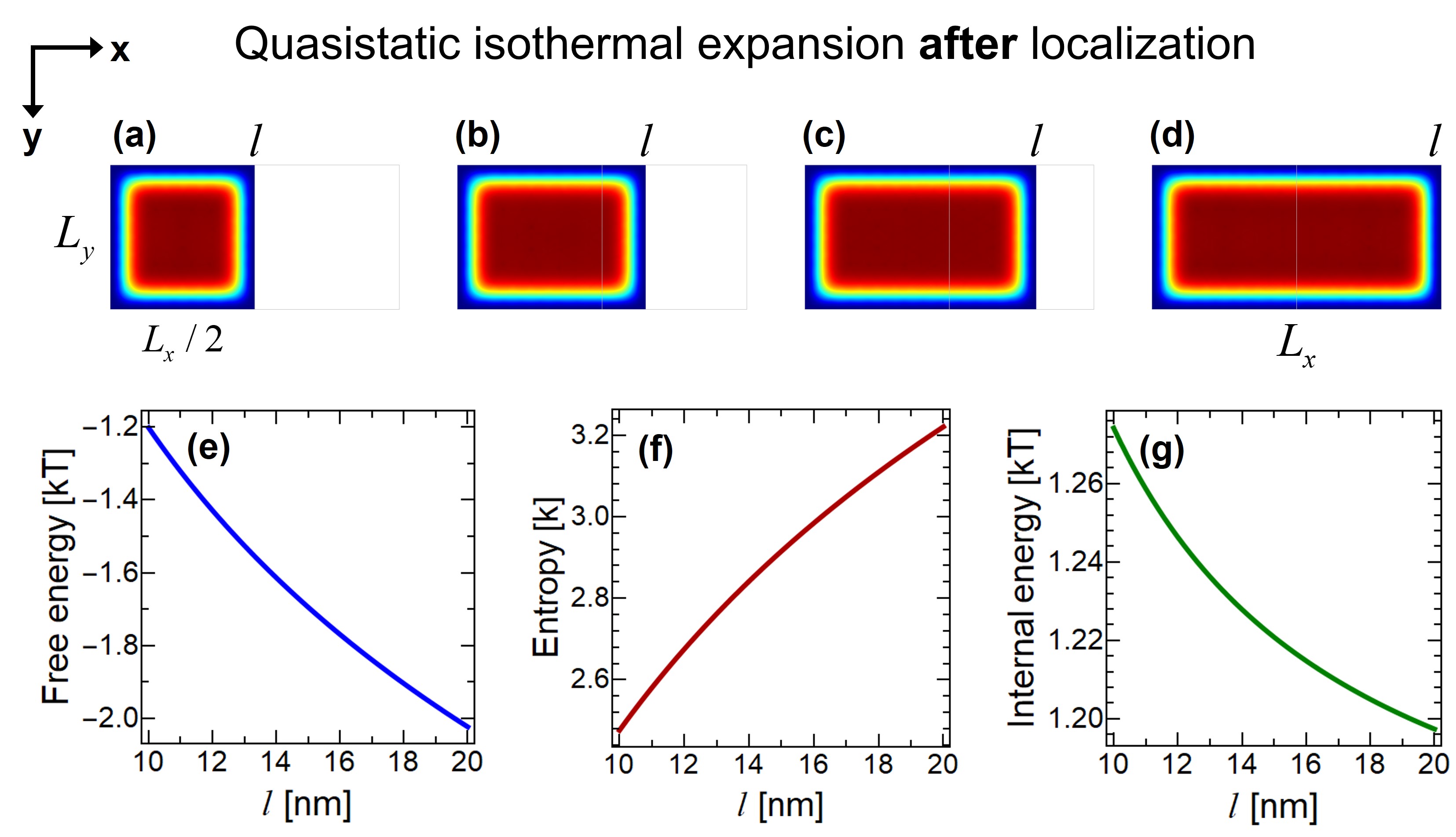}
\caption{Simulation of quasistatic isothermal expansion process after the measurement and localization of the particle. (a) After the symmetric insertion of the partition, quantum measurement is performed and the particle is localized at the left compartment. (b) Partition expands to the position of $l=14$nm, which is caused by the pressure exerted on the partition by the localized particle. (c) Expansion of the partition at $l=17$nm. (d) Partition expands to the right boundary of the container, $l=20$nm. Initial situation before the insertion is recovered. Variation of (e) Helmholtz free energy, (f) entropy and (g) internal energy with respect to partition's lateral position $l$ in nm's.}
\label{fig:pic6}
\end{figure}

\subsection{Step IV: Removal}

For the completion of the cycle, partition should be removed and the initial step (I) should be recovered. Since the wavefunction is exactly zero at the boundaries, removal of the insertion does not change any thermodynamic property of the system ($F_{\text{IV}}=F_{\text{I}}$, $S_{\text{IV}}=S_{\text{I}}$ and $U_{\text{IV}}=U_{\text{I}}$) and therefore can be done without any work and heat exchanges, as well as without a change in internal energy ($W_{\text{rem}}=0$, $Q_{\text{rem}}=0$ and $\Delta U_{\text{rem}}=0$). In single-particle Szilard engine with symmetric partition removal process is trivial. On the other hand, removal work can be finite in multi-particle case or container with finite-potential well boundaries.

Changes in thermodynamic quantities at each and every component of the quantum Szilard engine for each process are summarized in Table I. The first and second columns denote the components of the engine and differences in thermodynamic properties respectively. Rest of the columns represent the amounts of changes (given by the equations in Sec. III) in the thermodynamic properties for the steps of the cycle. It should be noted that work exchanges in the free energy row at $\mathcal{S}$ add up to zero, indicating that the extractable work is zero in a quantum Szilard engine. Furthermore, each row in $\mathcal{S}$ gives exactly zero and shows the cyclic nature of the processes. All columns add up to zero indicating that all processes in the engine are internally reversible. However, the measurement process is externally irreversible because of the nature of localization process, which relies on the collapse of the wavefunction. In this regard, localization of particle in a quantum Szilard engine brings quantum mechanics and thermodynamics into the same footing, where each of them plays their part respectively as irreversibility and heat dissipation. Demonstration of a purely quantum process obeying the Landauer's principle may have far-reaching implications and consequences in quantum-information thermodynamics, in addition to allowing the existence of demonless quantum Szilard engines.

\begin{table}[t]
\centering
\caption{Changes of free energy, entropy and internal energy in the system $\mathcal{S}$, device $\mathcal{D}$ and bath $\mathcal{B}$, for insertion (I), measurement (II), expansion (III) and removal (IV) processes.}
\label{table1}
\def\arraystretch{1.3}
\setlength{\tabcolsep}{0.5em}
\begin{tabular}{cccccc}
\hline
& 	                         & I & II & III & IV  \\ \hline
\multirow{3}{*}{$\mathcal{S}$}
&$\Delta F$ & +$W_{\text{ins}}$ & +$W_{\text{msr}}$ & $-W_{\text{msr}}$ $-W_{\text{ins}}$ & 0 \\
&$\Delta S$ & $-Q_{\text{ins}}$ & $-Q_{\text{msr}}$ & +$Q_{\text{msr}}$+$Q_{\text{ins}}$ & 0 \\
&$\Delta U$ & +$\Delta U_{\text{ins}}$ & 0 & $-\Delta U_{\text{ins}}$ & 0 \\
\hline
\multirow{3}{*}{$\mathcal{D}$}
&$\Delta F$ & 0 & $-W_{\text{msr}}$ & 0 & 0 \\
&$\Delta S$ & 0 & +$Q_{\text{msr}}$ & 0 & 0 \\
&$\Delta U$ & 0 & 0 & 0 & 0 \\
\hline
\multirow{3}{*}{$\mathcal{B}$}
&$\Delta F$ & $-W_{\text{ins}}$ & 0 & +$W_{\text{msr}}$+$W_{\text{ins}}$ & 0 \\
&$\Delta S$ & +$Q_{\text{ins}}$ & 0 & $-Q_{\text{msr}}$ $-Q_{\text{ins}}$ & 0 \\
&$\Delta U$ & $-\Delta U_{\text{ins}}$ & 0 & +$\Delta U_{\text{ins}}$ & 0 \\
\hline
\end{tabular}
\end{table}

\section{Conclusions and Future Work} 

We have presented a quantum Szilard engine in the absence of a Maxwell's demon. We've showed that even though there is no explicit information processing in demonless setups, measurement still plays a role and localization of the particle at one side is the crucial step for Szilard engine's operation. Although Maxwell's demon and Szilard engine problems are used interchangeably in literature because of the emphasis on erasure process, they can also be considered as separate in demonless quantum Szilard engine setups. Our work shows the role of Landauer's principle in demonless quantum Szilard engines, as we have pointed out, even if the acquisition of which-side information is irrelevant for the operation of a demonless Szilard engine, heat dissipation still takes place by another logically irreversible process, that is localization by quantum measurement. Therefore, Landauer's principle can save the second law when quantum-mechanical picture is taken into account in demonless Szilard engines.

In comparison with the erasure explanation, the main difference between the heat dissipation due to information erasure and due to localization by measurement is in the former "left or right" state goes back to a reference state whereas in the latter "left and right" state goes to the outcome state. Nevertheless, the localization by quantum measurement can also be interpreted by a mutual information exchange between the system and the measuring device, regardless of whether the device operates as an information-processing demon or not.

In this work, rather than concerning with the physical realizations of the Szilard engine, we have focused more on the abstract, conceptual version of it, to resolve some of the apparent problems in the case of absence of Maxwell's demon. We argue that erasure takes place at least implicitly in all demonless setups inherently because of the rectifying nature of the work extraction process, where two-way information has to be converted (rectified) into one-way information. For any kind of work extraction process, some kind of rectification has to be done albeit implicitly.

We've analyzed a demonless quantum Szilard engine explicitly and revealed that the localization holds the key along with Landauer's principle to save the second law and presents a complementary resolution of the quantum version of Szilard's paradox. Quantum mechanics was required for the justification of the third law of thermodynamics. Now, we've shown in this article that, it also saves the second law, suggesting that quantum mechanics has strong ties in the foundations of thermodynamics and information theory. Exploration of some implications and consequences of the particular role of Landauer's principle in localization might be significant to understand more about this link.

Work and heat exchanges and changes in energy for 1D and 2D Szilard boxes are the same and it can be generalized into n-dimension straightforwardly. But note that this won't be the case anymore, when one wants to use Fermi-Dirac or Bose-Einstein statistics. Magnitudes of quantum confinement effects will also differ in case of quantum statistics is used. Moreover, quantum degeneracy of the system (through chemical potential) will also play a significant role. Asymmetric insertion of the partition and multi-particle cases are also important to address from the point of view of quantum confinement effects. For instance, removal process might not be trivial anymore in the multi-particle case. Adiabatic and sudden processes rather than quasistatic ones may add interesting features to the original problem. Extension of this work into these cases is also possible. We believe our results on quantum Szilard engine without Maxwell's demon contribute to the understanding the link between information, quantum mechanics and thermodynamics.

\begin{acknowledgments}
We thank Robert Alicki and Saar Rahav for their helpful comments on the manuscript and Raam Uzdin for useful discussions. 
\end{acknowledgments}

\appendix*\section{Analytical expressions for work, heat and energy exchanges} 
Quantum confinement effects in thermodynamics can be understood and explained using quantum boundary layer framework\cite{qbl,qshe}. In addition to providing a physical understanding to the origins of quantum size and shape effects in confined systems, (e.g. here the quantum Szilard box), it allows one to obtain analytical expressions for many quantum-statistical systems with very high accuracy without the requirement of solving the Schr\"{o}dinger equation. Quantum boundary layer thickness for a confined ideal gas obeying Maxwell-Boltzmann distribution is exactly one-fourth of the thermal de Broglie wavelength, 
\begin{equation}
\delta=\frac{\lambda_{th}}{4}=\frac{h}{4\sqrt{2\pi mkT}},
\end{equation}
where $h$ is the Planck's constant. This thickness $\delta$ is derived from the thermal probability density distribution of a particle confined in 1D domain and contains both thermal and quantum natures of the particle(s) in an approximate and effective way, in some ways similar to the density matrix formalism. Under quantum boundary layer framework, we obtain insertion work analytically as
\begin{equation}
\begin{split}
W_{\text{ins}}^a=kT\ln\frac{L-2\delta}{L/2-2\delta}-kT\ln 2,
\end{split}
\end{equation}
where $a$ superscript denotes that the expression is analytical. Similarly, the change in internal energy during insertion process is found as
\begin{equation}
\begin{split}
\Delta U_{\text{ins}}^a=\frac{1}{2}kT\left(\frac{L/2}{L/2-2\delta}-\frac{L}{L-2\delta}\right).
\end{split}
\end{equation}
One can obtain analytical expression also for heat exchange during the insertion just by $Q_{\text{ins}}^a=\Delta U_{\text{ins}}^a-W_{\text{ins}}^a$. Note that work (and heat) exchanges during the measurement process are just $kT\ln 2$ (and $-kT\ln 2$) and during the expansion process thermodynamic exchange expressions contain the insertion terms as well, in addition to $kT\ln 2$. Therefore, by using Eqs. A.2 and A.3, work, heat and energy exchanges during all processes of the quantum Szilard engine cycle can be obtained analytically. As is seen, thermodynamic expressions can be determined just by considering the confinement geometry and quantum boundary layer $\delta$. In this regard, a quantum Szilard engine represents one of the most concrete examples showing the success of the quantum boundary layer concept.
\bibliography{sziref}

\begin{thebibliography}{10}
\bibitem{comsol}
All numerical simulations are implemented in {COMSOL}
  {M}ultiphysics\textsuperscript{\textregistered} software.
\end{thebibliography}
\bibliographystyle{unsrt}
\end{document}